\documentclass[twocolumn,aps,prd,showpacs,nofootinbib,superscriptaddress,preprintnumbers,floatfix,10pt]{revtex4-1}

\usepackage[english]{babel} 
\usepackage[utf8]{inputenc}
\usepackage[T1]{fontenc}
\usepackage{lmodern}

\usepackage{amsmath}
\usepackage{amsfonts}
\usepackage{amssymb}
\usepackage{slashed}

\usepackage{epsfig}
\usepackage{graphicx}
\usepackage{color}
\usepackage{epstopdf}

\usepackage{hyperref}

\newcommand{\I}{\mathrm{i}}
\newcommand{\E}{\mathrm{e}}

\newcommand{\STr}{\text{STr}}

\newcommand{\Nf}{N_{\mathrm{f}}}
\newcommand{\pt}{\partial_t}
\newcommand{\psib}{\bar{\psi}}

\newcommand{\UL}{U_{\Lambda}}

\newcommand{\mL}{m_\Lambda^2}
\newcommand{\lL}{\lambda_{2,\Lambda}}
\newcommand{\mH}{m_{\text{H}}}
\newcommand{\mtop}{m_{\text{t}}}

\newcommand{\htop}{h_{\mathrm{t}}}

\newcommand{\rmB}{\mathrm{B}}
\newcommand{\rmF}{\mathrm{F}}

\newcommand{\UMF}{U^{\mathrm{MF}}_{\mathrm{eff}}}

\newcommand{\UEMF}{U^{\mathrm{EMF}}_{\mathrm{eff}}}

\begin{document}

\preprint{}

\title{Nonpolynomial Higgs interactions and vacuum stability}

\author{Ren\'{e} Sondenheimer}
\email{rene.sondenheimer@uni-jena.de}
\affiliation{Theoretisch-Physikalisches Institut, Friedrich-Schiller-Universit\"at Jena, Max-Wien-Platz 1, D-07743 Jena, Germany}

\begin{abstract}
The possible violation of the conventional lower Higgs mass stability bound by the discovered Higgs boson has far reaching consequences within particle physics and cosmology.
We discuss the possibility that nonpolynomial bare interactions seeded at some high-momentum scale can considerably diminish the lower Higgs mass bound without introducing a metastability in the Higgs effective potential. For this, we classify various deformations of the usual quartic bare potential regarding their impact on stable IR physics.
We perform the analysis in a large $\Nf$ expansion, addressing the convergence of the obtained results by taking $1/\Nf$ corrections into account as well. 
%new
In addition, we investigate the renormalization group flow of the scalar potential on a nonperturbative level.
Within these approximations, we are able to identify bare potentials that lead to Higgs masses below stability mass bounds obtained from finite-order polynomial bare interactions without introducing a metastability in the effective potential.
\end{abstract}


\maketitle

\section{Introduction}
\label{intro}

The Higgs boson was the long term missing piece for the experimental confirmation of the standard model of particle physics. It took almost $20$ years from the commencement of construction of the LHC until the Higgs discovery in 2012 \cite{Aad:2012tfa,Chatrchyan:2012xdj}. The theoretical computation on mass bounds for the Higgs has an even longer history dating back to the 1970's. From renormalization group arguments it was known that the mass of the Higgs has to be in a finite infrared (IR) window for a given ultraviolet (UV) cutoff scale $\Lambda$ of the standard model \cite{Krive:1976sg,Maiani:1977cg,Krasnikov:1978pu,Politzer:1978ic,Hung:1979dn,Linde:1979ny,Cabibbo:1979ay,Lindner:1985uk,Kuti:1987nr,Wetterich:1987az,Lindner:1988ww,Sher:1988mj,Arnold:1989cb,Arnold:1991cv,Ford:1992mv,Sher:1993mf,Altarelli:1994rb,Espinosa:1995se,Hambye:1996wb,Isidori:2001bm,Fodor:2007fn,Einhorn:2007rv,Ellis:2009tp,Djouadi:2009nu,Holthausen:2011aa}. 
The current measurements of the standard model parameters, most prominently the top mass but also the value of the strong coupling constant or the masses of the electroweak gauge bosons, indicate that the mass of the Higgs violates the lower Higgs mass bound within the standard model for large $\Lambda$. This fact would render the effective Higgs potential metastable if it is assumed that the renormalization group running (RG) of the standard model parameters is only dominated by perturbatively renormalizable operators \cite{Buttazzo:2013uya,Gabrielli:2013hma,Bednyakov:2015sca,Iacobellis:2016eof,Chigusa:2017dux}.

The situation might change once degrees of freedom beyond the standard model enter the RG flow of the Higgs potential. These might stabilize the potential \cite{Hung:1995in,Casas:2000mn,Burgess:2001tj} or even compound the stability issue \cite{Loebbert:2015eea}. Thus, Higgs mass bounds can also be used to constrain parameters in different scenarios beyond the standard model and have been computed in various standard-model extensions \cite{Espinosa:1991gr,Casas:1994qy,Casas:1996aq,Bergerhoff:1999jj,Espinosa:2007qp,ArkaniHamed:2008ym,Kadastik:2011aa,EliasMiro:2011aa,Degrassi:2012ry,Lebedev:2012sy,Bezrukov:2012sa,EliasMiro:2012ay,Anchordoqui:2012fq,Lebedev:2012zw,Chen:2012faa,Masina:2012tz,Bezrukov:2014ina,Herranen:2014cua,Herranen:2015ima,Moss:2015gua,Okada:2015lia,Espinosa:2015zoa,Das:2015nwk,Das:2016zue,Abe:2016irv,Ema:2016ehh,Fischer:2016rsh,Branchina:2016bws,Bounakis:2017fkv}.

In the spirit of effective field theories, the yet unknown degrees of freedom beyond the standard model can be parametrized by higher-dimensional operators in order to perform model-independent analyses. These higher-dimensional operators are generically generated by the underlying structure of the standard model and can influence the RG running in various ways.

For instance, the impact of a bare $\lambda_{3}\phi^{6}$ coupling at the cutoff scale can diminish the lower Higgs mass bound in Higgs-Yukawa models mimicking the Higgs-fermion sector of the standard model \cite{Gies:2013fua,Gies:2014xha,Chu:2015nha}.
Incorporating also the influence of the gauge bosons on the RG running, it can be shown that Higgs masses $1$ GeV below the conventional stability bound at the Planck scale are still compatible with stable Higgs potentials \cite{Eichhorn:2015kea}.  A simple RG mechanism explains this fact. While the impact of the RG irrelevant coupling $\lambda_{3}$ on the other couplings rapidly dies out according to Wilsons arguments in the vicinity of the Gau{\ss}ian fixed-point, this operator is able to stabilize the Higgs potential in the deep UV even if the quartic Higgs coupling drops below zero. Thus, a possible instability scale can be shifted towards larger scales, leading to a diminishing of the lower Higgs mass bound.

As the conventional stability bound is usually associated with a vanishing quartic Higgs coupling at some UV scale, it is useful to introduce a new lower consistency bound for the mass of the Higgs once higher-dimensional operators are permitted within the bare action. 
The latter can be defined by the lowest possible Higgs mass given by a specified generalization of the bare action which has a Higgs potential equipped with a unique minimum during the entire RG flow. In particular this leads to the consistency condition that the potential is bounded from below to obtain a well-defined partition function of the theory.

The simple example of adding a $\phi^{6}$ term to the bare potential demonstrates that generalizations of the bare action can weaken the stability problem. 
In fact, the instability scale of the standard-model Higgs potential at $10^{10}\dots 10^{11}$ GeV can be shifted by at least one order of magnitude by this strategy \cite{Eichhorn:2015kea}.
Besides the simple extension of the bare action by polynomial Higgs self-interactions, also the impact of other polynomial generalizations of the bare interactions has been tested, e.g., in the Yukawa sector, confirming these results \cite{Jakovac:2015kka,Gies:2017zwf}.
However, this shift cannot be realized for an arbitrarily large amount of scales, as the running of polynomial higher-dimensional couplings is dominated by their power-counting behavior and thus they can contribute only for a comparatively small RG time to the flow. 
Due to this argument, this statement will likely hold for any class of polynomial bare interactions.

However, the existence of an absolute lower consistency bound is an involved minimization problem in theory space spanned by all possible bare potentials.
Especially, the impact and RG running of nonpolynomial bare interactions on the Higgs mass stability issue is not explored in detail, so far. 
A further relaxation of the lower consistency bound might be possible, if a suitable nonpolynomial bare potential exists such that the RG flow of the Higgs potential stays for a sufficiently long RG time away from its power counting behavior and the usual RG arguments in the vicinity of the Gau{\ss}ian fixed-point can be circumvented.

The aim of this work is to cast a first glance on possible modifications of the effective Higgs potential and a diminishing of the lower Higgs mass bound by nonpolynomial bare potentials. 
For this task, we use a systematic $1/\Nf$ expansion to investigate the properties of the effective potential.
In Sec.~\ref{sec:MF}, we will introduce a toy model to concentrate on the Higgs-top interactions as the top Yukawa coupling is the driving force for the stability problem. 
After defining the theory, we will compute the effective potential for the scalar field within a mean-field analysis which represents the leading order contribution of the large $\Nf$ expansion.
In order to improve our analysis, we take $1/\Nf$ corrections in an extended mean-field analysis into account in Sec.~\ref{sec:EMF}. 
In particular, we give constructive suggestions of possible nonpolynomial bare interactions that lead to Higgs masses substantially below the lower consistency mass bound of any bare action spanned by a set of finite-order polynomials. While it is straightforward to obtain such bare potentials in the mean-field approximation, the consideration of scalar fluctuations can spoil the convergence properties of the large $\Nf$ expansion. Nonetheless, the inclusion of scalar fluctuations offers new mechanisms to diminish the lower mass bound at the same time. 
%new
Inspired by these results, we check how RG improvement alters the results. In particular many nonpolynomial classes show unsatisfactory convergence properties within the $1/\Nf$ expansion. In Sec.~\ref{sec:Flow}, we investigate the RG running of the full scalar potential on a functional level for specific examples and reveal properties of the underlying UV physics to obtain a stable effective potential.
We finally conclude and give an outlook in Sec.~\ref{sec:Con}.

\section{Mean-field analysis}
\label{sec:MF}

As the large top mass dominates the RG flow of the Higgs quartic coupling and is responsible for the fact that it becomes negative at large RG scales, we will focus on a simple Higgs-Yukawa model mimicking the Higgs-top sector of the standard model in the following. This toy model has proven useful to investigate the occurrence of Higgs mass bounds in the literature also on a nonperturbative level \cite{Gies:2013fua,Holland:2003jr,Holland:2004sd,Branchina:2005tu,Fodor:2007fn}, see \cite{Gies:2017ajd} for a brief review. The classical, Euclidean action of the model is given by
\begin{align}
 S = \int_{x} 
 \left[ \frac{1}{2}(\partial_\mu \phi)^2 + U(\phi) +  \psib \I \slashed{\partial} \psi + \I \htop \phi \psib\psi \right]. 
 \label{eq:bareaction}
\end{align}
Demanding that the scalar potential is invariant under a $\mathbb{Z}_{2}$ symmetry, $U(\phi)=U(-\phi)$, the action exhibits a discrete chiral symmetry, $\phi \to -\phi$, $\psi \to \E^{\I \frac{\pi}{2} \gamma_5} \psi$, and $\psib \to \psib \E^{\I \frac{\pi}{2} \gamma_5}$, which mimics the properties of the electroweak symmetry group within this toy model. Particularly the Dirac fermion, which represents the top quark,  can acquire a mass term only due to spontaneous symmetry breaking.

In order to get a first, simple approximation of the effective potential which is obtained after all fluctuations are integrated out, let us investigate the fermionic partition function of this model. As the fermions appear only as a bilinear in the action, we can integrate them out yielding the standard fermion determinant of a Yukawa theory. We perform this computation in Euclidean spacetime for convenience.
\begin{align}
 Z_{\rmF} = \int_{\Lambda} \mathcal{D}\psi \mathcal{D}\psib \E^{- S[\phi,\psi,\psib]} =  \E^{- S_{\rmB}[\phi]} \det(\I \slashed{\partial} + \I\htop\phi),
\end{align}
where $S_{\rmB}$ is the purely bosonic part of the classical action $S$ defined in Eq.~\eqref{eq:bareaction}. 
The UV cutoff scale at the functional integral indicates that we formulate this theory in the spirit of an effective field theory with an intrinsic finite cutoff belonging to the definition of the model. 
Technically, this scale can be viewed as a UV regularization.
However, it is also associated to a physical scale.
Below that scale the considered theory can be formulated in terms of a viable quantum field theory to describe certain aspects of a physical system.  For larger scales, the model loses its validity and has to be replaced by a more fundamental theory. 
As the standard model likely has to be defined with such an upper validity scale and is only an effective description of nature, we explicitly introduce a finite but arbitrary UV cutoff scale in our toy model.
%For this reason, we keep the cutoff dependence of  throughout all computations. 
%In this spirit, our strategy is similar to the four-Fermi theory of the beta decay. The process can be described in terms of perturbatively nonrenormalizable operators at least below the electroweak scale which sets a natural cutoff for a suitable description of the beta decay in terms of a four-Fermi theory. These operators are generated by the underlying structure which is the electroweak standard model in this case. As we can only give upper bounds on the cutoff scale of the standard model and do not know which type of interactions will be generated by the underlying structure, we test in the following whether the stability scale can be shifted towards larger scales once nonpolynomial interactions are considered.}

In order to extract the effective potential at an one-loop level, we consider a homogenous mean-field for the scalar field, $\phi(x) = \mathrm{const.}$.
This is sufficient as the fermionic determinant already corresponds to a loop integration. Deviations from this homogeneous field configuration contribute only at a higher loop level. 
Therefore, we obtain for the fermionic induced effective mean-field potential
\begin{align}
 \UMF(\phi) = U_{\Lambda}(\phi) - \frac{1}{2\Omega} \ln\frac{\det_{\Lambda}(-\partial^{2} + \htop^{2} \phi^{2})}{\det_{\Lambda}(-\partial^{2})},
\end{align}
where we have chosen a normalization of the generating functional that the fermion-induced effective action is normalized to the zero field limit and $\Omega$ denotes the spacetime volume. Moreover, we used the hermiticity property of $\gamma_{5}$, i.e., $\I\slashed{\partial}$ is isospectral to $-\I\slashed{\partial}$.

The ratio of the functional determinants can be evaluated straightforwardly once a suitable regularization procedure is chosen. We use in the following a linear regulator family as is often used in the context of functional RG equations \cite{Litim:2000ci,Litim:2001up}, in particular in the context of Higgs mass bounds \cite{Gies:2013fua,Gies:2014xha,Eichhorn:2014qka,Eichhorn:2015kea,Jakovac:2015kka,Borchardt:2016xju,Gies:2017zwf,Gies:2017ajd}.
Thus, we use this type of regulator for reasons of convenience to directly compare with these studies. Moreover, functional flows or nonperturbative lattice simulations along the lines of \cite{Gerhold:2007yb,Gerhold:2007gx,Gerhold:2009ub,Gerhold:2010bh,Gerhold:2010wv,Bulava:2013ep,Chu:2015nha} will be needed to further improve the following large $\Nf$ analysis as we will demonstrate in the next sections. 
We emphasize, that the following conclusions remain the same for other type of regulators like a sharp momentum cutoff, zeta-function regularization, or various classes of mass dependent regularization schemes \cite{Gies:2014xha}.

The effective mean-field potential can be computed analytically for the linear regulator and reads
\begin{align}
 \UMF = U_{\Lambda} - \frac{1}{16\pi^{2}} \left[ \Lambda^{2} \htop^{2} \phi^{2} - \htop^{4}\phi^{4} \ln\left( 1 + \frac{\Lambda^{2}}{\htop^{2}\phi^{2}} \right) \right]. 
\end{align}
This approximation of the effective potential becomes exact in the strict limit $\Nf \to \infty$, assuming the model exhibits $\Nf$ copies of Dirac fermions. In the context of Higgs mass bounds, the simple mean-field approximation has turned out to be a remarkable good approximation already for $\Nf=1$ in case the top fluctuations dominate the RG flow of the scalar couplings, i.e., for the conventional lower mass bound. The mean-field lower bound deviates only slightly from a nonperturbative investigation of the stability bound including threshold effects, RG improvement, as well as a full functional flow of the scalar potential \cite{Gies:2013fua,Borchardt:2016xju}.

\subsection{Higgs mass consistency bound for polynomial bare potentials}
The main advantage of this simple-minded approximation is that the effective mass of the scalar particle can be analytically computed. It can be expressed as a function of the UV cutoff of the model as well as of the bare parameters encoded in the bare potential $\UL$ \cite{Gies:2013fua},
\begin{align}
 \mH^{2} &= {\UMF}''(v) \notag \\
 &= \frac{\mtop^{4}}{4\pi^{2}\, v^{2}} \left[ 2 \ln\left( 1+ \frac{\Lambda^{2}}{\mtop^{2}} \right) - \frac{3\Lambda^{4} + 2\mtop^{2}\Lambda^{2}}{(\Lambda^{2} + \mtop^{2})^2} \right] \notag \\
 &\quad+ \UL''(v) - \frac{1}{v}\UL'(v),
 \label{eq:HiggsMassMF}
\end{align}
where $v$ is the nontrivial minimum of the effective potential of the scalar field ${\UMF}'(v) = 0$, given by the Fermi scale in the standard model. We exchanged the bare Yukawa coupling by the top mass parameter as we fix this coupling in the deep IR by a suitable renormalization condition which is given by $\mtop = \htop v$ for our simple approximation. 
Again, this is an oversimplification of the complex RG flow of the standard model but sufficient for our qualitativ investigation at the moment.
Even though we consider only a toy model here, we choose $\mtop = 173$~GeV and $v = 246$ GeV in order to make contact with standard-model physics in the following.

Assuming that the bare potential at the cutoff scale is given by only perturbatively renormalizable operators, i.e., $\UL = \frac{\mL}{2}\phi^{2} + \frac{\lL}{8}\phi^{4}$, we get,
\begin{align}
 \mH^{2} 
 &= \frac{\mtop^{4}}{4\pi^{2}\, v^{2}} \left[ 2 \ln\left( 1+ \frac{\Lambda^{2}}{\mtop^{2}} \right) - \frac{3\Lambda^{4} + 2\mtop^{2}\Lambda^{2}}{(\Lambda^{2} + \mtop^{2})^{2}} \right]
 \notag \\ &\quad + \lL v^{2},
 \label{eq:HiggsMassQuartic}
\end{align}
yielding a mass which is a monotonically increasing function of the bare quartic coupling $\lL$ for a given cutoff $\Lambda$ and fixed top mass $\mtop$. Thus, we obtain a natural lower mass bound for the Higgs, $\mathrm{min}\, \mH = \mH(\lL = 0)$, for the class of quartic bare potentials, for which the Higgs mass is entirely build up from top fluctuations. Lower Higgs masses cannot be meaningfully obtained in this Higgs-Yukawa model, as already the bare potential would be unbounded from below for negative bare quartic couplings. Hence, the effective potential would suffer from an instability as well. 
This conclusion is a direct consequence from the fact that the asymptotic behavior of the potential cannot be altered by the RG running as can be seen from the properties of exact RG flow equations \cite{Wetterich:1992yh}, for instance.

However, as long as the underlying structure of the standard model is unknown, other interactions beyond the power counting renormalizable operators cannot be excluded at the cutoff scale. Currently, no experiment is able to put constraints on these higher-dimensional operators. The simplest possible extension of the quartic bare potential is by other polynomial interactions at the cutoff scale,
\begin{align}
\UL = \frac{\mL}{2}\phi^{2} + \frac{\lL}{8} \phi^{4} + \frac{\lambda_{3,\Lambda}}{48 \Lambda^{2}} \phi^{6} + \frac{\lambda_{4,\Lambda}}{4!2^{4} \Lambda^{4}} \phi^{8} + \cdots.
\label{eq:BarePotPoly}
\end{align}
Including these operators in the computation of the Higgs mass, we obtain
\begin{align}
 \mH^{2} 
 &= \frac{\mtop^{4}}{4\pi^{2}\, v^{2}} \left[ 2 \ln\left( 1+ \frac{\Lambda^{2}}{\mtop^{2}} \right) - \frac{3\Lambda^{4} + 2\mtop^{2}\Lambda^{2}}{(\Lambda^{2} + \mtop^{2})^{2}} \right] \notag \\ 
&\quad + v^{2} \left[ \lL +  \frac{\lambda_{3,\Lambda}}{2}\frac{v^{2}}{\Lambda^{2}} + \frac{\lambda_{4,\Lambda}}{8}\frac{v^{4}}{\Lambda^{4}} + \cdots \right].
\label{eq:HiggsMassPoly}
\end{align}
The contribution from the RG irrelevant couplings $\lambda_{n\geq 3,\Lambda}$ to the effective mass of the Higgs field is suppressed by suitable powers of the cutoff $\Lambda$ as one would expect from a dimensional analysis in the vicinity of the Gau{\ss}ian fixed-point. Thus, for a sufficient large separation of the electroweak scale from the scale of new physics, the IR observables are almost independent of these modifications of the bare action and are far beyond the current precision measurements.

Even though the higher-dimensional operators do not have a direct impact on the observable IR Higgs mass, they modify the stability considerations and thus have an indirect impact on the position of the lower stability bound. At this point it is important to keep in mind that the stability mass bound does not contain only information about the IR physics but also of the UV embedding of the standard model. In the presence of positive $\lambda_{n\geq 3,\Lambda}$ a negative bare quartic coupling can be permitted in the UV, as the higher-order couplings can potentially stabilize the scalar potential without introducing a meta- or instability on all RG scales.

Let us exemplify this by a generalization of the bare potential by a simple $\lambda_{3}\phi^{6}$ operator along the line of \cite{Gies:2013fua,Gies:2014xha,Borchardt:2016xju}. 
For quartic bare potentials, Eq.~\eqref{eq:HiggsMassQuartic} can be viewed from two perspectives once the mass of the Higgs is known. We can either fix the quartic coupling by the mass of the scalar particle for a given cutoff or we are able to compute the scale of maximal UV extent of the model which is determined by the lower mass bound $\lL=0$.
If a Higgs mass of $125$ GeV is required, the scale of maximal UV extent is given by $\Lambda_{\phi^{4}} \sim 10^{7}$ GeV within our Higgs-top toy model for a top mass of $173$ GeV.
To push the cutoff scale even further, negative values of the bare quartic coupling have to be chosen which induce an instability in the bare potential as well as in the effective potential. 
This problem can be circumvented once a $\lambda_{3,\Lambda} \phi^{6}$ operator is allowed. 
The requirement of a bare potential that is bounded from below translates into a positive $\lambda_{3,\Lambda}$ coupling. 
Having a negative quartic coupling, the lower mass bound is indeed diminished as the contribution from the positive $\lambda_{3,\Lambda}$ to the effective Higgs mass is highly suppressed by the cutoff, see Eq.~\eqref{eq:HiggsMassPoly}, which leads effectively to a larger cutoff for a fixed Higgs mass. 
Besides implications for the Higgs mass an additional $\phi^{6}$ operator affects also tunneling rates in case a second minimum is present \cite{Branchina:2013jra,Branchina:2014rva,Branchina:2014usa,Lalak:2014qua,Bentivegna:2017qry}, see also \cite{Branchina:2015nda} for a specific beyond the standard model scenario, or the electroweak phase transition \cite{Akerlund:2015fya,Chu:2017vmc}.

Unfortunately, the instability scale cannot be arbitrarily shifted by this simple generalization. Suppose $\lambda_{3,\Lambda} = 3$. For this value, the bare quartic coupling can safely be diminished until it reaches $\lL = -0.065$. For smaller $\lL$ the bare potential can be stable with a unique minimum at vanishing field amplitude, however, the effective potential develops a second nontrivial minimum rendering the effective potential metastable due to the interplay of the nontrivial structure of the bare potential $\UL$ and the top fluctuation induced part of the effective potential \cite{Borchardt:2016xju}. While for a quartic bare potential the extremal condition of the effective minimum ${\UMF}'=0$ has only one nontrivial $\mathbb{Z}_{2}$-symmetric solution, the richer polynomial structure allows for more solutions in the generalized case.
Thus, the metastability arises for different reasons than the previous stability problem for quartic bare potentials.
Nonetheless, even for the seeming small value of $\lL=-0.065$ the cutoff scale can be shifted by an order of magnitude to $\Lambda \sim 10^{8}$ GeV.

This simple example demonstrates how irrelevant interactions can weaken the stability issue. Nonetheless, the large gap between the instability scale in the standard model and the Planck scale can unlikely be bridged by polynomial interactions at the cutoff scale. Of course, it is possible to add more terms beyond the $\phi^{6}$ generalization. However, for these type of finite-order polynomial bare interactions, the second minimum in the effective potential beyond the Fermi minimum is usually at the order of the cutoff scale $\phi_{\mathrm{min}}/\Lambda \sim \mathcal{O}(1)$ and generically developed by a first order phase transition during the RG flow if not already present in the bare potential for sufficiently large absolute values for $\lL$. As these higher-dimensional operators are even more strongly suppressed by the cutoff scale, and the corresponding couplings $\lambda_{n}$ die out faster, any finite-order approximation of the bare potential in terms of polynomial interactions will not be able to prevent a metastability in the effective Higgs potential for a sufficiently light Higgs.

Of course, an exception could be given by rather exotic finite-order polynomials that have a large higher-order coupling, $\lambda_{n}\gg 1$. For instance, the scale of maximal UV extent can be pushed to $\Lambda \sim 10^{9}$ GeV if $\lambda_{3,\Lambda} = 100$ for $\mH = 125$ GeV. 
As a rule of thumb within this mean-field approximation, a coupling $\lambda_{3,\Lambda} \sim \mathcal{O}\big(\Lambda^{2}/(10\Lambda_{\phi^{4}})^{2}\big)$ is required to stabilize the scalar potential where $\Lambda_{\phi^{4}}$ is the instability scale if only power counting renormalizable operators are considered in the bare action. 
Nevertheless, this type of solution comes with a grain of salt. Albeit it cannot be ruled out a priori, it is very unlikely that the underlying structure of the standard model generates a finite-order polynomial potential for the scalar field that singles out one (or a few) dimensionless coupling, say $\lambda_{3,\Lambda}$ for simplicity, which is orders of magnitude larger than the other coupling constants.

From the Wilsonian view point every interaction term that is compatible with the field content and the symmetries of the model will be present at the cutoff scale. Especially the scalar potential is an arbitrary function of the field amplitude $\phi$ as long as it respects the $\mathbb{Z}_{2}$ symmetry. Restricting the discussion to a quartic bare potential or a bare potential with $\phi^{6}$ term assumes implicitly that the bare potential is expandable in a meaningful Taylor series at the origin. 
In the first instance, it is reasonable to assume that the dimensionless higher-order couplings $\lambda_{n,\Lambda}$ of this Taylor series are of order one, also to guarantee a suitable radius of convergence to obtain trustable results within a finite-order approximation. The situation might change once an infinite series is considered with increasing higher-order coupling strength. For this, a full functional analysis as well as appropriate resummation is required.

\subsection{Higgs mass consistency bound for nonpolynomial bare potentials}
In case of a finite-order Taylor-like bare potential, we have seen that a new lower consistency bound can be formulated. This bound is a few GeV below the conventional stability mass bound which is derived for power counting renormalizable operators but still guarantees a unique minimum of the potential at all RG scales. 
However, it is only able to push the conventional mass bound by one order of magnitude towards larger scales.
Also, polynomial generalizations in other sectors of the bare action, e.g., by generalized Yukawa interactions $h(\phi^{2})\phi\psib\psi$ \cite{Jakovac:2015kka,Gies:2017zwf}, seem to not further diminish this lower mass bound. 
Thus, this bound might be universal for any bare action with polynomial interactions where the higher-order dimensionless bare couplings are of order $\mathcal{O}(1)$.

In order to further diminish the lower Higgs mass consistency bound, we now focus on nonpolynomial bare interactions. 
A variety of possibly viable extensions regarding the stability issue might exist in the infinite dimensional theory space of all possible bare potentials. 
Minimizing the lower consistency bound is thus an intricate problem and clearly beyond the scope of this work. 
We will rather classify the implications of different nonpolynomial structures within the bare potential on the stability issue and the IR physics and present constructive examples that diminish the polynomial lower bound without introducing a metastability in the effective potential in the mean-field approximation and beyond.

In particular, we investigate three different cases. Bare potentials which can not be expanded in a Taylor series at vanishing field amplitude, potentials with a finite radius of convergence, and potentials which can be written in a power series with infinite radius of convergence.
Some of these potentials might be motivated by underlying physics that can be described in the context of a quantum field theory, like Coleman-Weinberg type potentials which arise by integrating out heavy degrees of freedom. By contrast, the underlying structure of the standard model does not necessarily be explainable by yet known methods and techniques. For this reason, we do not want to restrict to a specific scenario.

\subsubsection{Bare potentials with vanishing radius of convergence}
The lower mass bound is essentially built up from the logarithmic term in Eq.~\eqref{eq:HiggsMassMF} induced by top fluctuations. 
As a first example, let's try to weaken this impact by modifying the standard $\phi^{4}$ potential by a logarithmic structure that will eat up the fermion fluctuations, 
\begin{align}
 \UL = \frac{\mL}{2}\phi^{2} + \frac{\lL}{8}\phi^{4} - a \phi^{4} \ln \left(1+ \frac{\Lambda^{2}}{b\phi^{2}} \right),
 \label{eq:BarePotLog}
\end{align}
with positive constants $a$ and $b$. Note, that this bare potential and also the effective mean-field potential is bounded from below if and only if $\lL > 0$.
For further convenience, we choose $a=b^{2}/(16\pi^{2})$ as this is sufficient for our following purpose. 
In this case it is straightforward to see that parameter regions exist that can diminish the lower Higgs mass bound drastically without introducing an instability. The simplest example is given by the choice $b= h^{2}$. The logarithmic modification of the quartic bare potential exactly cancels the top fluctuation induced part in the mean-field potential. Thus, the effective mean-field potential only has a simple $\phi^{4}$ form and is stable for positive $\lL$ which is anyhow required for a stable bare potential. 
The Higgs mass can then be freely adjusted according to the precise value of the quartic coupling for any value of the cutoff scale.

Also for other values of $b$, the impact of the fermionic fluctuations can be significantly weaken, depending on the ratio $b/h^{2}$. 
Inserting the bare potential \eqref{eq:BarePotLog} into the mean-field approximation of the Higgs mass \eqref{eq:HiggsMassMF}, the lowest possible value of $b$ can be determined by the consistency constraint $\lL>0$ for a given cutoff and Higgs mass. For instance, for $b>0.36$ the cutoff scale of our toy model can be pushed by at least five orders of magnitude compared to quartic bare potentials towards $\Lambda = 10^{12}$~GeV for $\mH = 125$ GeV without introducing a metastability or instability in the scalar potential. For smaller values of $b$, a negative bare quartic coupling is needed to obtain the desired Higgs mass, rendering the potential unstable. 
Larger values of $b$ allow for a further increase of $\Lambda$.
Similar analyses can also be performed for $a \neq b^{2}/(16\pi^{2})$, of course, where large regions of the parameter space regarding $a$ and $b$ exist which diminish the lower bound considerably once this particular logarithmic modification of the bare potential is permitted.

Besides this specific logarithmic extension of the bare potential, we tested a variety of other functions. 
The obvious difference between the $\ln$-type bare potential and polynomial generalizations is the singular structure of the potential \eqref{eq:BarePotLog} at the origin, yielding a potential which cannot meaningfully expanded in a polynomial around the minimum at the origin as $\lL \sim \lim_{\phi \to 0} \ln(1/\phi^{2})$ and ${\lambda_{n\geq 3,\Lambda} \sim \lim_{\phi\to 0} 1/\phi^{2n-4}}$.

\subsubsection{Bare potentials with finite radius of convergence}
Let us now investigate whether bare potentials with a finite radius of convergence can solve the stability problem. For this task, we slightly modify our previous example \eqref{eq:BarePotLog} by a mass-type coupling parameter $\mu$,
\begin{align}
 \UL = \frac{\mL}{2}\phi^{2} + \frac{\lL}{8}\phi^{4} - a \phi^{4} \ln \left(1+ \frac{\Lambda^{2}}{\mu^{2}\Lambda^{2} + b\phi^{2}} \right).
 \label{eq:BarePotLogFinite}
\end{align}
Expanding the potential \eqref{eq:BarePotLogFinite} in a power series around its minimum at $\phi=0$, we obtain a radius of convergence in units of the cutoff scale $\Lambda$ which is given by $\mu/\sqrt{b}$. For simplicity, we choose $b=1$ in the following.
We use this specific function again for purely illustrative purposes. Similar conclusions hold for other functions which have a Taylor series expansion at the origin with a finite radius of convergence like $a \phi^{4} \ln(1+b\phi^{2}/\Lambda^{2})$, $a \phi^{4} \arctan{(b\phi^{2}/\Lambda^{2})}$, or $a\phi^{4}/(1+b\phi^{2}/\Lambda^{2})$.

Regarding the stability issue, we observe the following. We are able to diminish the lower mass bound even below the consistency bound of generalized polynomial bare potentials if a suitable value of $\mu$ is chosen. In order to shift the cutoff by $n$ orders of magnitude from the $\phi^{4}$ instability scale $\Lambda = 10^{n} \Lambda_{\phi^{4}} \simeq 10^{7+n}$ GeV, the parameter $\mu$ has to be of the order $\mathcal{O}(10^{-n})$ or smaller. This implies that the nonpolynomial structure of Eq.~\eqref{eq:BarePotLogFinite} is able to solve the stability problem only if the radius of convergence is close to or smaller than the instability scale $\Lambda_{\phi^{4}}$ as one would naively expect.

From a conventional perspective one might be tempted to argue that new physics has to show up below the scale $\Lambda_{\phi^{4}}$, based on these results. For instance, structures as they appear in the potential \eqref{eq:BarePotLogFinite} might be generated from a heavy massive bosonic particle which couples directly to the Higgs field and has a mass given by $\mu\Lambda$. As only for $\mu\Lambda \lesssim \Lambda_{\phi^{4}}$ the potential is stabilized, the occurrence of new physics is below the instability scale, solving the stability problem trivially. However, we would like to emphasize at this point that this has not necessarily to be the case.

From a more conservative point of view, Nature might be only described by the degrees of freedom and symmetries of the standard model up to scales $\Lambda \gg \Lambda_{\phi^{4}}$, if nonperturbative effects in terms of nonpolynomial structures in the bare potential are present and dominate the RG flow above a certain scale given by $\mu\Lambda \lesssim \Lambda_{\phi^{4}}$. 
In this case, the Higgs potential can be meaningfully described in terms of a polynomial series at small field amplitudes, $\phi < \mu\Lambda$, especially near the electroweak scale, implying that a perturbative description suffice to explain current collider data. Above the scale $\mu\Lambda$ nonperturbative effects seeded by the bare action at some high scale $\Lambda$ may render the effective potential stable without introducing new degrees of freedom or new particles below the cutoff scale.

One might be worried about the fact that a seemingly unnatural small value for $\mu$ has to be generated at the cutoff scale to obtain a sufficiently large separation between the cutoff and the instability scale. However, the parameter $\mu$ is not associated to a specific coupling as usually occurs in a perturbative analysis but rather contributes to the specific properties of a full coupling functional in terms of the potential~\eqref{eq:BarePotLogFinite} and a functional investigation for all field amplitudes is needed to capture the entire nonperturbative effects. In that sense we formulate no constraint on this parameter. 
It rather classifies to which subspace the potential belongs in theory space.
In that sense, the specific example for the bare potential in Eq.~\eqref{eq:BarePotLogFinite} can be understood as a placeholder for any potential with an analogous structure. It is merely chosen for an illustrative example in terms of elementary functions.

\subsubsection{Bare potentials with infinite radius of convergence}
Besides the two considered examples in Eq.~\eqref{eq:BarePotLog} and Eq.~\eqref{eq:BarePotLogFinite} representing bare potentials which have not a well-defined polynomial expansion at the minimum or a finite radius of convergence respectively, also a third possibility can lead to the desired properties which we already have sketched at the end of the previous subsection. Suppose the underlying theory of the standard model generates an infinite polynomial series with an infinite radius of convergence but sufficiently strong higher-order interaction terms. Then, the Taylor approximation of the potential converges for every field amplitude but with a slow rate of convergence such that very high truncation orders are needed to capture the relevant properties.

For this type of bare potentials, we use a simple exponential function for illustration,
\begin{align}
 \UL = \frac{\mL}{2}\phi^{2} + \frac{\lL}{8}\phi^{4} + a \phi^{4}\, \E^{\frac{b\phi^2}{2\Lambda^{2}} }.
 \label{eq:BarePotExp}
\end{align}
In case $b$ (and $a$) are of order $\mathcal{O}(1)$ or smaller, only a few terms in a Taylor approximation are needed to properly investigate the properties of the effective potential regarding the instability issue and we fall back into the discussion below Eq.~\eqref{eq:BarePotPoly} as the bare higher-dimensional couplings $\lambda_{n,\Lambda}$ are of order one. 
The situation changes if $b\gg 1$. In this case, the higher-order couplings grow according to $\lambda_{n,\Lambda} \sim b^{n-2}$ for $n>2$ until the factorial $n!$ in the denominator of the series coefficients of the exponential function takes over ensuring the convergence properties of the Taylor series. Depending on the precise value of $b$, several terms have to be considered within the polynomial approximation and especially the 'low-dimensional' coupling constants $\lambda_{3}$, $\lambda_{4}$, $\cdots$ become large. 
However, this is not problematic as the full series can be added up to an exponential function with large $b$ by construction within our example.

In order to diminish the lower bound by this strategy, a sufficient large $b$ has to be chosen such that the occurrence of a second minimum at large field amplitudes $\phi_{\mathrm{min}}\sim \Lambda$ driven by a negative $\lL$ is suppressed but still small enough that the impact of the new contributions do not alter the small field behavior of the plain $\phi^{4}$ structure. Otherwise the lower mass bound would increase due to the strong coupling of the higher-order operators. Our rule of thumb derived for the $\phi^{6}$ class of bare parameters is already a good indication for the specific example given by Eq.~\eqref{eq:BarePotExp} as the potential can be expressed in terms of a power series where $\lambda_{3,\Lambda} \sim b$. In order to shift the cutoff scale $n$ orders of magnitude away from the $\phi^{4}$ instability scale, $\Lambda = 10^{n}\Lambda_{\phi^{4}}$, $b$ has to be of the order $\mathcal{O}\big(\Lambda^{2}/(10\Lambda_{\phi^{4}})^{2}\big)$. This might imply rather large values for $b$ but again, we deal here with a full coupling functional instead of an extension in terms of an additional single coupling. In the sense the parameter $\mu$ was used for the bare potential~\eqref{eq:BarePotLogFinite} to classify the nonpolynomial effects that lead to a finite radius of convergence, $b$ can be used to pick an example of the class of potentials with a specific rate of convergence towards the full function. 
Then, a large value $b$ signals that a sufficiently slow rate of convergence is required.

Aside from this example with a rather large parameter, also potentials can be constructed with parameters of order one for the sake of complexity regarding the functional dependence on the field amplitude. For instance the cutoff scale can be pushed towards $10^{9}$ GeV in our toy model for a bare potential given by,
\begin{align}
 \UL = \frac{\mL}{2}\phi^{2} + \frac{\lL}{8}\phi^{4} + a \phi^{4}\, \E^{b\, \E^{\frac{c \phi^2}{2\Lambda^{2}} }}
\end{align}
for $a=1$, $b=c=2$ or to $\Lambda = 10^{10}$ GeV for $a=1$, $b=c=4.75$. Similarly higher values of the scale of maximal UV extent can be approached, e.g., by replacing the exponential by $\mathrm{exp}\big(b\, \mathrm{exp}(c\, \mathrm{exp}(d \phi^{2}/\Lambda^{2}))\big)$, we can achieve $\Lambda=10^{11}$ GeV for $b=c=d=1.7$.

To briefly summarize, two strategies can be used to weaken or even solve the stability problem of the standard model Higgs sector in terms of generalized Higgs interactions at least in the large $\Nf$ limit.
First, the nonpolynomial structure has no impact on the shape of the effective potential near the electroweak scale. Then, a negative quartic coupling is needed to diminish the lower mass bound and the nonpolynomial interactions have to compensate the occurrence of a second minimum at large field values near the cutoff scale driven by the negative quartic coupling. The last class of potentials with a sufficiently slow convergence rate belongs to this case.
Second, the deviation from the $\phi^{4}$ structure can directly affect the effective quartic coupling at the electroweak scale and thus the Higgs mass. In case it suppresses the contribution coming from the top quark, the lower mass bound can be diminished as well without introducing a metastability in the effective potential. For our examples of $\ln$-type modifications, we ensured that the large field behavior is governed by a positive bare quartic coupling which avoids the occurrence of a second minimum.

\section{Extended mean-field analysis}
\label{sec:EMF}

So far, we only used a simple mean-field approximation in order to calculate the effective potential, which is the first contribution in a large $\Nf$ expansion. As long as the bosonic sector is only weakly coupled and the top Yukawa coupling dominates the RG flow, this approximation has turned out to be useful even for small $\Nf$ not only qualitatively but also to some extent on a quantitative level for the lower mass bound \cite{Gies:2013fua} as well as the effective potential \cite{Borchardt:2016xju}, at least for polynomial type bare interactions. To improve our  understanding of the nonpolynomial bare potentials, an improved calculation for the effective potential is mandatory as for some field amplitudes the system becomes strongly coupled and the validity of the mean-field approximation cannot be guaranteed.

An extended mean-field calculation is the next logical step as this approximation takes $1/\Nf$ corrections into account by including the scalar fluctuations on the same Gau{\ss}ian level as the fermionic fluctuations. The resulting determinant can be computed analytically for the class of linear regulator functions which we used in the previous section and the extended mean-field effective potential reads,
\begin{align}
 \UEMF = U_{\Lambda} &- \frac{\Nf}{16\pi^{2}} \left[ \Lambda^{2} \htop^{2} \phi^{2} - \htop^{4}\phi^{4} \ln\left( 1 + \frac{\Lambda^{2}}{\htop^{2}\phi^{2}} \right) \right] \notag \\
 & + .\frac{1}{64\pi^{2}} \left[ \Lambda^{2} {\UL''} -{\UL''}^{2} \ln\left( 1 + \frac{\Lambda^{2}}{\UL''} \right) \right]
 \label{eq:EMF}
\end{align}
where primes denote derivatives with respect to $\phi$ and we reinstated $\Nf$ merely as an ordering parameter of the calculation. For all quantitative statements, we use ${\Nf=1}$.

\subsubsection{Bare potentials with vanishing radius of convergence}
At first glance, the logarithmic extension of the quartic structure in Eq.~\eqref{eq:BarePotLog} seems as an appropriate extension. However, incorporating the scalar fluctuations to the renormalization process, we obtain a strong contribution from the curvature of the bare potential induced by the singular structure of the logarithm at the origin. Especially the quartic coupling defined at the electroweak scale, $\lambda_{2,\mathrm{eff}} = {\UEMF}^{(4)}(\phi=v)$, renormalizes with an unusual behavior as the polynomial bare couplings obtained from an expansion at the electroweak scale behave as $\lambda_{n,\Lambda} \sim \Lambda^{2n-4}/v^{2n-4}$ for $n>2$ and $\Lambda \gg v$. Therefore, we obtain the peculiar situation of a unique minimum at the electroweak scale but Higgs masses of the order of the cutoff scale within the extended mean-field approximation. Note that this result obviously does not diminish the lower mass bound but circumvent the upper triviality bound due to nonperturbative effects. Nonetheless, the upper bound cannot meaningfully be dealt with within the mean-field or extended mean-field approximation as RG improvement is mandatory for such a strongly coupled Higgs sector even in the simple case of quartic bare potentials. 

Whether a full nonperturbative RG investigation which includes RG improvement can wash out this strong renormalization at the electroweak scale, leading indeed to a diminishing of the lower bound, or circumvent the triviality arguments for the upper bound cannot be answered a priori. 
At this point, we are only able to conclude that the singular behavior of the bare potential~\eqref{eq:BarePotLog} spoils the convergence of the $1/\Nf$ expansion for a large scale separation between the cutoff and the electroweak scale and a full nonperturbative RG investigation is required to make a definite statement.%
%%% new
We perform such an investigation in Sec.~\ref{sec:Flow}.

Of course, this problem does not occur for small $\Lambda$ only a few scales above the electroweak scale, e.g., {$\Lambda = 10$ TeV}, with a suitable value $a<1$. 
However, already polynomial generalizations with $\lambda_{3,\Lambda} \sim \mathcal{O}(1)$ can considerably diminish the lower Higgs mass bound for small cutoff scales.

Instead of the nonpolynomial structure of the bare potential $\UL$ itself, there also is the possibility that the scalar fluctuations induced by the curvature of the nonpolynomial bare potential $\UL''$ compensate the renormalization coming from the top for a negative $a$ with $|a| \ll 1$. 
This is only possible if the dimensionless parameter $a$ compensates the large contribution $\Lambda^{2}/v^{2}$ coming from the strong curvature of the bare potential near the origin, i.e., $a \sim v^{2}/\Lambda^{2}$. For instance, we obtain a stable effective potential  with $\mH = 125$ GeV for $\Lambda = 10^{10}$ GeV, if $a = -5.6 \times 10^{-15}$. Nevertheless, the reliability of this result is questionable due to the qualitative difference between the mean-field and extended mean-field results caused by the large effects of the scalar fluctuations as well as RG improvement is still missing in this simple computation.

\subsubsection{Bare potentials with finite radius of convergence}
In a similar way the $\ln$-type example with finite radius of convergence, Eq.~\eqref{eq:BarePotLogFinite}, does not show the desired convergence properties.
First, we observe that the contribution induced from the scalar fluctuations to the renormalized effective quartic coupling and thus to the Higgs mass is $\sim \mu^{-2}$ as can be seen by a straightforward computation,  
\begin{align}
 \mH^{2}
 &= \lL v^{2} - 8a \ln (\mu^{-2}) v^{2} + \frac{\Nf\, \mtop^{4}}{4\pi^{2}\, v^{2}} \left[ 2 \ln\left( \frac{\Lambda^{2}}{\mtop^{2}} \right) - 3 \right]  \notag \\ 
&\quad  + \frac{15ab}{4\pi^{2}} \frac{1}{\mu^{2}} v^{2} + \mathcal{O}\left(\frac{v^{2}}{\Lambda^{2}}\right),
\label{eq:HiggsMassLog}
\end{align}
for $v\ll \Lambda$, $\mu \ll 1$, and $bv^{2} \ll \mu\Lambda$, where we have separated the contribution from the scalar fluctuations in the second line.
The first line contains the contribution from the top fluctuations (last term ${\sim}\Nf$) as well as the curvature of the bare potential at the electroweak scale in the first two terms which gets renormalized by the fluctuations, i.e., the first line on the right-hand side represents the mean-field result. 
For the mean-field case a sufficiently small $\mu$ was needed to compensate the top contributions and to ensure that the radius of convergence drops below $\Lambda_{\phi^{4}}$ such that the nonperturbative effects can stabilize the potential for large field values. 
The scalar fluctuations included in the extended mean-field approximation can thwart the diminishing for too small $\mu$.
Thus, we have to first answer the question whether parameters exist such that these two contrary effects can be balanced to solve the stability problem, before we turn towards the convergence properties of this specific example in the $1/\Nf$ expansion.

Choosing negative $a$, a critical value $\mu_{\mathrm{cr}}$ can be found that minimizes the Higgs mass for a given $\Lambda$. For $\mu < \mu_{\mathrm{cr}}$ the radius of convergence shrinks which strengthens the nonperturbative effects, leading to larger Higgs masses and spoiling the convergence of the $1/\Nf$ expansion. For $\mu > \mu_{\mathrm{cr}}$, the radius of convergence becomes larger, implying that the nonpolynomial structure cannot prevent the effective potential from becoming metastable. 
Nonetheless, the lower bound obtained by this strategy can be below the lower consistency bound for the class of generalized polynomial bare potentials.

However, convergence regarding the large $\Nf$ expansion cannot be expected since the diminishing mechanisms are qualitatively different between the mean-field and extended mean-field approximation. 
The nonpolynomial deformation of the bare potential contributing to a modification of the bare quartic coupling at the electroweak scale, see first two terms on the right-hand side of Eq.~\eqref{eq:HiggsMassLog}, and the curvature of the bare potential determining the scalar fluctuations (second line of Eq.~\eqref{eq:HiggsMassLog}) come with opposite sign. Thus, a change in the sign of $a$ is necessary to obtain stable bare potentials with a Higgs mass below the conventional stability mass bound by going from mean-field to extended mean-field, as in the previous case. 
This leads to the fact, that every set of parameters for the bare potential~\eqref{eq:BarePotLogFinite} that solves the stability problem in the mean-field approximation does not provide a solution for the extended mean-field case and vice versa.

This problem might be circumvented by potentials of this class for which the bare contribution and the contribution induced by scalar fluctuations contribute with the same sign, e.g., for $\arctan(\phi^{2})$ or $\ln(1+\phi^{2})$. However, we were not able to find a set of parameters for these potentials that diminish the lower mass bound considerably below the lower consistency bound of the $\phi^{6}$ class within the extended mean-field approximation.

\subsubsection{Bare potentials with infinite radius of convergence}

The scalar fluctuations can spoil the convergence properties of the large $\Nf$ expansion also for the bare potential~\eqref{eq:BarePotExp} belonging to the class of potentials which can be expanded in a polynomial for arbitrarily large field amplitude but sufficiently slow convergence rate. Nevertheless, there are regions in parameter space for this example in which the extended mean-field approximation show merely moderate deviations from the mean-field results.

\begin{figure}[t!]
\centering
\includegraphics[width=0.95\columnwidth]{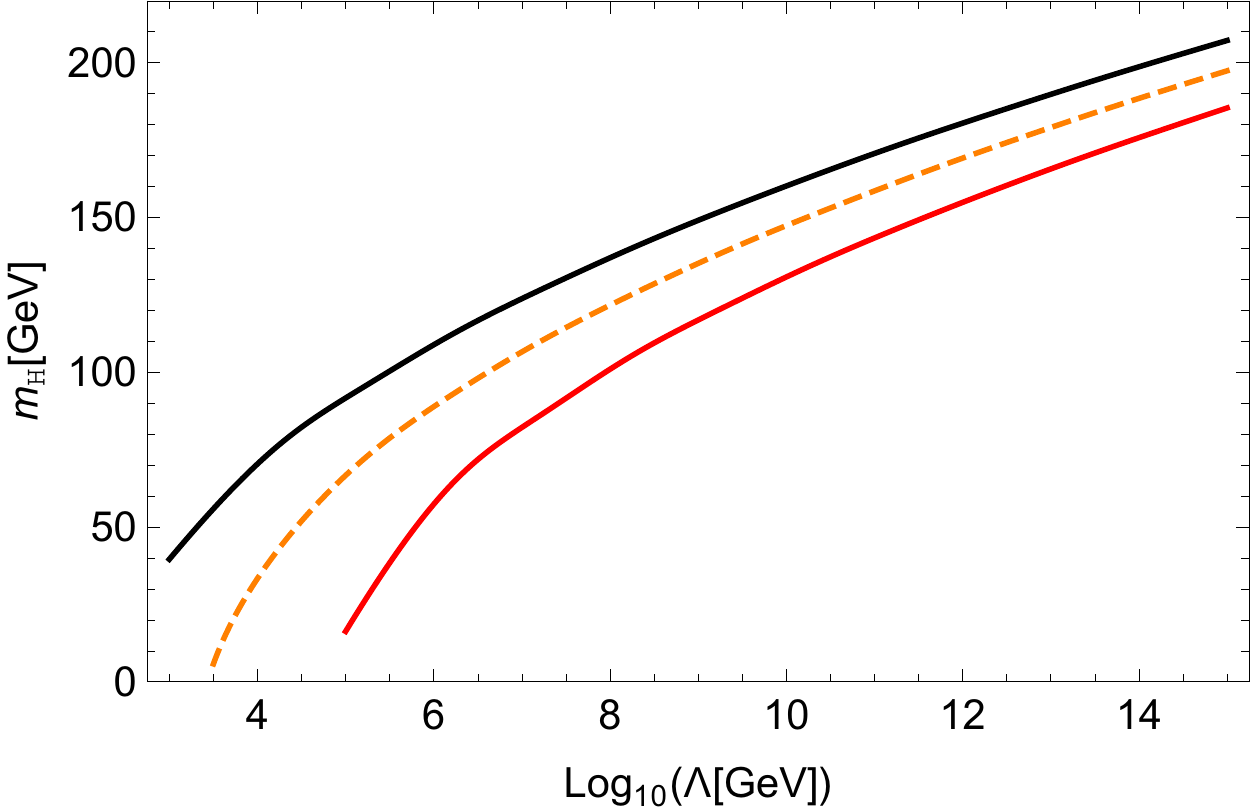}
\caption{Comparison of Higgs mass consistency bounds for different bare potentials. The black solid curve belongs to the conventional lower stability mass bound for quartic bare potentials. The orange dashed line is obtained from the lowest possible Higgs masses for the class of $\phi^{6}$ bare potentials with a unique minimum for the bare as well as the effective potential. The red solid line depicts the lower consistency mass bound for bare potentials given by Eq.~\eqref{eq:BarePotExp} for $a=10^{-4}$ and ${\lL=-0.18}$.}
\label{fig:HiggsMass}
\end{figure}

The contribution to the Higgs mass induced by the scalar fluctuations is ${\sim} b$ for the class of bare potentials modified by an exponential function.
The larger $b$, i.e., slower rates of convergence, the stronger the system is coupled such that no convergence of the results can be expected by the current simple approximations of the effective potential and RG improvement is required again. 
By contrast, the occurring metastability cannot be prevented for too large values of the cutoff for too small $b$. 
Following the same strategy as in the previous case of bare potentials with a finite radius of convergence, we are able to determine an upper critical value for $b$ which balance both effects. 
The lower mass bound determined by $b_{\mathrm{cr}}$ is depicted for $\lL=-0.18$ and $a=10^{-4}$ in Fig.~\ref{fig:HiggsMass} as red solid line. For comparison, we plotted also the conventional lower stability mass bound for $\phi^{4}$ bare potentials as black solid line and the lower consistency bound for the $\phi^{6}$ generalization with $\lambda_{3,\Lambda}=3$ as orange dashed line. 
Comparing the conventional lower mass bound to the consistency mass bound of the exponential bare potential, the scale of maximal UV extent can be shifted by almost three orders of magnitude for this specific example.

In order to compare this lower mass bound to the mean-field results, we fix the parameters $a$ and $b$ of the bare potential but vary $\lL$ until the effective potential becomes metastable within the mean-field approximation. Comparing the obtained values for the masses within both approximations, we observe a deviation of the Higgs mass by at most $10\%$ for the region of interest $\Lambda > 10^{6}$ GeV. 
This moderate deviation between the mean-field and extended mean-field Higgs mass can be traced back to the specific properties of potentials with an infinite radius of convergence but small convergence rate. The parameters $a$ and $b$ appear in a particular combination such that the small field behavior of the scalar potential is governed by the usual power-counting renormalizable structure while for field amplitudes close to the cutoff the generation of a second minimum is avoided by the strong couplings $\lambda_{3}$, $\lambda_{4}$, $\cdots$. 
%%% new
In order to trust these results beyond the large $\Nf$ expansion, we perform a full nonperturbative RG calculation in Sec.~\ref{sec:Flow}.
%%%%%

\subsubsection{Beyond elementary functions}

After the promising results of the mean-field calculation, the extended mean-field results do not favor a scenario with a rather simple nonpolynomial generalization of the bare potential such that the scale of maximal UV extent can be shifted towards the Planck scale. % apart from the the last discussed class of bare potentials. 
Although, suitable bare potentials can be constructed leading to stable extended mean-field approximations for the effective potential, most of them called for RG improvement to obtain a reliable result.
At least, we were able to construct an example that further diminishes the lower consistency bound by a few GeV without spoiling a possible convergence of the $1/\Nf$ expansion for the class of potentials given by an infinite polynomial series but sufficient slow rate of convergence.

However, we would like to emphasize, that we only investigated bare potentials which were expressed in terms of elementary functions, so far. 
The space of all allowed bare potentials is much larger. For instance, it is possible to numerically construct a bare potential that can circumvent the stability problem by rethinking Eq.~\eqref{eq:EMF}. 
This equation can be viewed as a nonlinear second order differential equation to obtain a suitable bare potential once the effective potential is fixed. The two integration constants can be fixed by demanding that the solution respects the $\mathbb{Z}_{2}$ symmetry of the model, $\UL'(0)=0$, and by choosing a convenient value for the in our case unimportant offset of the potential, e.g., $\UL(0) = 0$. 
This yields a unique solution for the bare potential once the effective potential is specified.  By this strategy it can be tested, which stable IR physics can be extended up to sufficient high energy scales, in case a solution to this nonlinear differential equation exist.

\begin{figure}[t!]
\centering
\includegraphics[width=0.95\columnwidth]{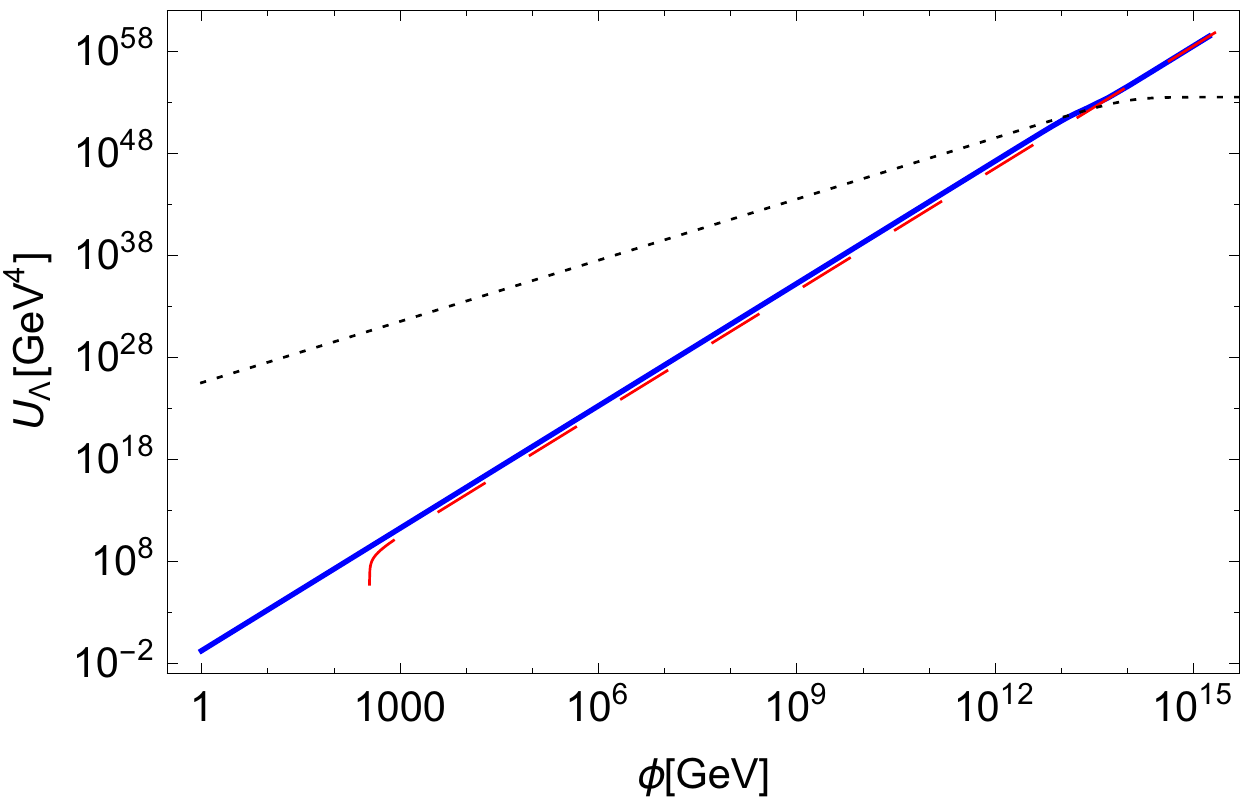}
\caption{Numerical solution of Eq.~\eqref{eq:EMF} for the bare potential $\UL$ (blue solid line) for $\Lambda = 10^{14}$ GeV. The IR physics is governed by a quartic Higgs potential (red dashed line) by construction. The black dotted curve shows the contribution of the scalar fluctuation induced part as well as the absolute value of the fermion induced part. As the difference of both contributions are hardly visible by eye in this double logarithmic plot, they appear as one line, indicating that both contributions almost compensate each other.}
\label{fig:EMFPot}
\end{figure}

A numerical solution of this problem is depicted in Fig.~\ref{fig:EMFPot} where the bare potential is plotted as blue solid line. For simplicity, we have assumed that the effective potential (red dashed line) is only given by a stable $\phi^{4}$ potential equipped with a minimum at the electroweak scale and a Higgs mass of $125$ GeV. The scale of new physics is set to $10^{14}$ GeV. Albeit the solution for the bare potential looks rather trivial at logarithmic scales, it has a variety of noteworthy properties. 
The contribution coming from the scalar fluctuations to the effective potential (second line of Eq.~\eqref{eq:EMF}, depicted as black dotted line in Fig.~\ref{fig:EMFPot}) is almost identical to the absolute value of the fermion determinant for field values larger than the electroweak scale. Thus, we observe a dynamical cancellation between both contributions such that no second minimum is generated at large field values and the effective potential is stable.

For large field amplitudes $\phi \sim 100 \Lambda$ the differential equation becomes stiff, making it challenging to go to arbitrarily large amplitudes. Nevertheless, already at scales slightly above the cutoff scale, the scalar as well as the top fluctuations approach constant values and thus do not modify the large field behavior which is given by $\phi^{4}$ by construction. For scales below $\Lambda$, we observe slight deviations from the quartic structure being strong enough that the effective potential does not develop a second minimum but small enough near the origin such that the IR physics is not affected by this modification and a Higgs mass of $125$ GeV can be obtained.

Besides the example depicted in Fig.~\ref{fig:EMFPot}, we also investigated the construction of the bare potential via reverse engineering for other cutoff values as well as different stable, weakly coupled IR potentials. In all cases, the solutions behave in a similar way as described above. 
Thus, not the plain modification of the quartic structure accounts for the diminishing of the lower mass bound without introducing a metastability as was suggested in the mean-field approximation but the scalar fluctuations described by the curvature of the bare potential. 
In this case, the scalar fluctuations have to play a similar dominant role as the top fluctuations but are not given in terms of a single strong coupling constant though induced by the nonpolynomial deformation from the quartic structure. 
This behavior was also seen for the $\ln$-type modifications above. 

Let us finally highlight, that the example depicted in Fig.~\ref{fig:EMFPot} is also below the lower mass bound for the exponential bare potential plotted as red solid line in Fig.~\ref{fig:HiggsMass}. Even though there is no convergence regarding the $1/\Nf$ expansion for most of the investigated generalizations, we are optimistic that the reverse engineering of the bare potential can also be used for a full nonpertrubative flow equation study in subsequent work.

\section{Nonperturbative RG flow of the scalar potential}
\label{sec:Flow}

In order to improve our results, a full nonperturbative RG study is required as most modifications of the potential include nonperturbative structures and effects. 
In particular it is important to verify whether the stabilizing effects will be washed out once RG improvement is included.
For this, the functional RG approach formulated in terms of the Wetterich equation \cite{Wetterich:1992yh} is an ideal tool. The Wetterich equation
\begin{align}
 \pt \Gamma_{k} = \frac{1}{2} \STr \Big[ \big(\Gamma_{k}^{2} + R_{k} \big)^{-1} \pt R_{k} \Big], \quad \pt = k \frac{\mathrm{d}}{\mathrm{d}k},
\end{align}
interpolates smoothly between the classical action defined at the cutoff scale $S = \Gamma_{k=\Lambda}$ and the full effective action $\Gamma = \Gamma_{k=0}$ via an IR cutoff $R_{k}$ and allows to investigate the strong coupling limit, threshold effects, and the RG evolution of a full coupling function depending on various mass scales. 
For instance, the flow equation for the dimensionless scalar potential ($u = k^{-d}U$) for the considered Yukawa model can be obtained by a systematic derivative expansion and reads,
\begin{align}
 \pt u &= -d\, u + \frac{1}{2}(d-2+\eta_{\phi})\phi u' \notag \\
 &\quad+ 4v_{d} \left[l_{0}^{(\rmB)d}\big(u'';\eta_{\phi}\big) - d_{\gamma} l_{0}^{(\rmF)d}\big(\phi^{2} h^{2};\eta_{\psi}\big) \right],
 \label{eq:flowPot}
\end{align}
where primes denote derivatives with respect to the scalar field $\phi$ and $\eta_{\phi}$ and $\eta_{\psi}$ are the anomalous dimensions of the scalar and fermion field, respectively. %The dimensionless potential is introduced via $u = k^{-d}U$ 
The threshold functions $l_{0}^{(\rmB/\rmF)d}$ encode the loop integration over bosonic and fermionic degrees of freedom. These can be performed analytically for the linear regulator family which we used in Sec.~\ref{sec:MF} and \ref{sec:EMF}. The threshold functions as well as the nonperturbative flow equations for the anomalous dimensions and the Yukawa coupling for the considered model can be found, e.g., in \cite{Gies:2017zwf}.

The flow equations for the quartic coupling, the mass parameter of the scalar field, or any other higher-dimensional scalar-self coupling can be extracted form Eq.~\eqref{eq:flowPot} via suitable projections. Moreover, also the RG flow of the entire scalar potential with nonpolynomial interactions can be addressed by solving this partial differential equation. Of course, this is rather time consuming compared to the functional investigation of the large $\Nf$ expansion because a numerically stable solution has to be obtained over many orders of magnitude regarding the RG scale $k$ as well as the field amplitude $\phi$ to separate the electroweak from the cutoff scale.

The large $\Nf$ expansion has shown that the class of polynomials with infinite radius of convergence exhibit promising properties to solve the stability issue. 
It is at least reasonable to expect that this type of diminishing is also present in the full flow for the following reason. 
Usually, the impact of the higher-dimensional coupling $\lambda_{3}$ on the quartic coupling $\lambda_{2}$ is washed out after a few RG scales as the RG running of $\lambda_{3}$ is governed by its power counting behavior. In case the running of $\lambda_{3}$ is driven by a large $\lambda_{4}$ for a sufficiently long RG time, the impact on $\lambda_{2}$ can be extended. The even faster die-out of $\lambda_{4}$ can be compensate by an even larger coupling $\lambda_{5}$ and so on. 
A similar mechanism can also be used to circumvent the triviality problem of the scalar sector in gauged-Higgs models which become asymptotically free \cite{Gies:2016kkk}.
Therefore, we restrict our following considerations mainly to this specific class. Nonetheless, as the higher-dimensional couplings behave as $b^{n}/n!$ for the exponential bare potential \eqref{eq:BarePotExp} the described mechanism can only bridge a finite (but possibly arbitrary) amount of scales as $b^{n}/n! \to 0$ for fixed $b$ and $n \to \infty$.

\begin{figure}[t!]
\centering
\includegraphics[width=0.95\columnwidth]{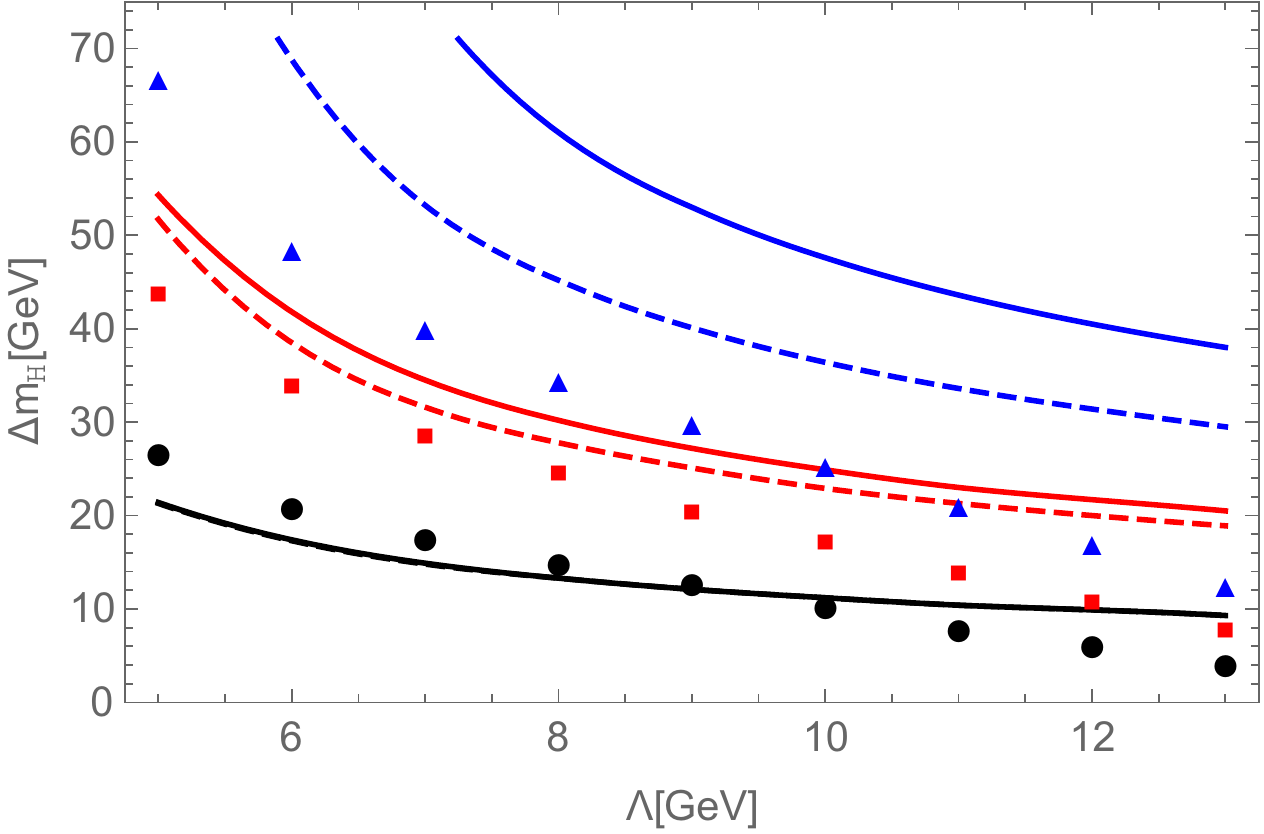}
\caption{Deviation of the lower Higgs mass consistency bound for exponential-type bare potentials from the lower stability bound for quartic bare potentials. The black curves correspond to $b=1/2$ while the red and blue curved are computed for $b=5$ and $b=50$, respectively. Dashed lines depict mean-field results, solid lines take $1/\Nf$ corrections into account, and the circles ($b=1/2$), squares ($b=5$), and triangle ($b=50$) show results of the full RG flow of the scalar potential including RG improvement.}
\label{fig:Comparison}
\end{figure}

A useful property of this class is that some of the characteristics of the full functional solution can be investigated in a polynomial truncation of the potential. 
A similar observation has been made for bare potentials with finite-order polynomials. Although, a polynomial projection on the flow of the potential covers only local information in field space, the radius of convergence at intermediate RG scales $k$ is usually still large enough to spot a potential metastability for polynomial-type bare potentials \cite{Borchardt:2016xju}. We also observe this behavior for exponential-type bare potentials for sufficiently high truncation orders. We examine this by comparing the solutions of the full flow, i.e., solving the partial differential equation \eqref{eq:flowPot}, to a finite polynomial approximation $\lambda_{n}\phi^{2n}$ up to $n=16$ for selected initial conditions.

According to our previous investigations on the stability issue for finite-order bare potentials, we have checked the convergence of our results for different truncations. These checks include improvements of the derivative expansion by comparing results of a local potential approximation to results which include scale-dependent wave function renormalizations. In addition we tested the stability of our results by including other higher-order operators from the Yukawa sector by allowing for a (polynomial) Yukawa potential $h(\phi^{2})$ during the RG flow. Technical details on such truncation test can be found in \cite{Gies:2013fua,Gies:2014xha,Borchardt:2016xju,Gies:2017zwf}.
For the class of exponential-type bare potentials, we observe satisfactory convergence properties even in the strong coupling limit similar to the previous results for polynomial bare potentials. 

Most importantly, we are able to find initial conditions for the flow equation which can considerably diminish the lower mass bound of quartic bare potentials as well as the lower consistency mass bound for finite-order bare potentials. Choosing Eq.~\eqref{eq:BarePotExp} as initial potential at the cutoff scale with $a=1$ and positive $b$, we are able to choose a negative quartic coupling as long as the potential is stabilized by the exponential modification. In qualitative agreement to the large $\Nf$ expansion, we observe that some critical $\lambda_{2,\Lambda}^{\mathrm{cr}}$ exist for fixed $b$ which defines a new lower consistency bound for this specific class of bare potentials. For $\lambda_{2,\Lambda} > \lambda_{2,\Lambda}^{\mathrm{cr}}$ the scalar potential is stable during the entire RG flow, while for $\lambda_{2,\Lambda} < \lambda_{2,\Lambda}^{\mathrm{cr}}$ a second minimum is generated due to the nontrivial interplay between the scalar and fermionic fluctuations. We plot the deviation of this new lower consistency bound from the lower stability bound for quartic bare potentials in Fig.~\ref{fig:Comparison} as black circles for $b=1/2$, red squares for $b=5$, as well as blue triangles for $b=50$. 

For comparison, we also plot the results obtained from the mean-field and the extended mean-field investigation as dashed and solid lines, respectively. For sufficiently small $b$ where the scalar potential is still in a regime which can be described with perturbative techniques or can be approximated by a finite polynomial including only a few terms, the deviation between mean-field and extended mean-field results (black curves) is hardly visible by eye. Likewise the deviation from the full flow equation study is small such that the large $\Nf$ approximation of the effective potential is a suitable tool to obtain a first glance on the IR properties in this regime. Once $b$ is increased, the solutions start to deviate on a quantitative level but at least the qualitative features can be captured by all different approximations. It seems that the extended mean-field results generically overshoot the impact of the scalar fluctuations. This is not surprising as RG improvement is missing in the $1/\Nf$ expansion. Thus the strongly coupled scalar fluctuations contribute over too many scales as only the bare propagators are used to integrate out modes. Their contribution is weakened in a full flow equation study as the large contributions from higher-dimensional couplings die out during the flow. Nonetheless, the impact of these nontrivial interactions modifies the flow of the potential in the UV in such a way that the scalar potential remains stable during the entire RG flow. 

The diminishing effect decreases for larger values of the cutoff like for the case of a finite-order polynomial modification of the bare interactions. 
Nonetheless, we would like to emphasize that we were able to demonstrate that the instability scale can be shifted by 3 orders of magnitude with the considered initial conditions up to $b=50$ and the difference of the resulting Higgs masses between the lower bounds is by a factor $2$-$3$ bigger for the exponential modification compared to any finite-order polynomial.
Going to even larger values of $b$ and thus lower Higgs masses is not a conceptual but numerical issue as it becomes challenging to compute a numerical stable solution in this case.

So far, we have only investigate the exponential function given in Eq.~\eqref{eq:BarePotExp} as a representative of a bare potential with infinite radius of convergence. 
However, it is not likely that the underlying physics of the standard model will solely generate an exponential modification of the standard quartic structure of the Higgs potential at the cutoff scale $\Lambda$. 
Nonetheless, we would like to emphasize that the results presented here will be similar for any potential which can be expanded in a Taylor series with sufficient slow rate of convergence.
In order to substantiate this conjecture, we perform the following tests.

First, we investigate variations of the plain exponential structure given in Eq.~\eqref{eq:BarePotExp}. Therefore, we add a fixed order monomial $\frac{c_{N}}{n!}\phi^{2N}$ to the exponential modification of the quartic Higgs potential. The results in the following do not alter if either the full functional flow of the bare potential or only a (sufficient high) finite-order approximation of the exponential function is studied. In case of a finite order polynomial approximation, we ordinarily choose $N$ to coincide with the highest order exponent but the results do not change if $N$ is smaller. Now, we crank up the coupling $c_{N}$ which serves as a measure for the departure of the exponential. As this test becomes numerically expensive for increasing $N$, we focus on $b=1/2$ as well as $b=5$ for $\Lambda=10^{6}, 10^{7}$, and $10^{8}$ GeV, and $b=50$ at $\Lambda=10^{7}$ GeV. We choose these cutoff values simply because the instability scale of this toy model is of order $\mathcal{O}(10^{7})$ GeV for a Higgs mass of $125$ GeV for the considered toy model. 

For all tests we find approximately the same pattern. The modification influences the low energy physics only if a certain critical order of magnitude of the coupling $c_{N}$ is approached. For instance, the stability of the Higgs potential and the IR Higgs mass is not altered as long as $c_{4}<10$. Once $c_{4}$ becomes $\mathcal{O}(10)$, we obtain a slight increase of the Higgs mass of $\mathcal{O}(0.1)$ GeV and a shift of a few GeV if $c_{4}$ is $\mathcal{O}(100)$. As long as the Higgs mass increases, the potential remains stable during the entire RG flow. For larger $N$ the maximal order of magnitude of the coupling increases. It can be estimated by $c_{N} \approx 10^{2.5N-8}$. As long as $c_{N}$ is smaller, the IR physics is altered by less than a GeV. Thus, we observe a certain flexibility of the UV potential around the exponential function.

Apart from this study, we have also checked that similar shifts of the lower Higgs mass bound are possible for other functional structures, e.g., by replacing the exponential by a $\cosh$ or a nested exponential structure like $\E^{\E^{\tilde{b} x^{2}/2}}$. 
For instance, the shift of the lower Higgs mass bound for $\tilde b \approx 0.2$ is roughly the same as in case of the exponential modification with $b=1/2$. 
As long as the lower order coefficients of the Taylor expansion of the investigated function are of the same size as in the exponential case, we find similar shifts of the Higgs mass consistency bounds without introducing a metastability in the Higgs potential. 

This fact can also be understood from the above mentioned point of view.
The stability issue of the Higgs potential and mass is mainly governed by the running of the quartic coupling for the class of bare potentials with infinite radius of convergence. This running is directly modified by $\lambda_{3}$ and (in the broken regime) $\lambda_{4}$. Higher-order couplings have only an indirect impact via the running of these two couplings. Thus, the lower order contributions of the expansion will have the dominant impact as long as higher order couplings do not become exorbitant large. Any function with a low order Taylor expansion similar to the exponential function will result in the same IR physics and therefore a similar shift of the mass bound. Thus, we view the exponential just as a representative of the class of functions which can be expanded in an infinite Taylor series with a certain rate of convergence.

\section{Conclusions and Outlook}
\label{sec:Con}

In this work, we addressed the impact of nonpolynomial bare interactions on the stability of the Higgs potential and the related lower Higgs mass consistency bound. 
We found that deviations from the usual polynomial interactions might have the possibility to circumvent the RG arguments which lead to a metastability of the Higgs potential at large field values. 
It was possible to construct various classes of bare potentials that lead to an considerably shift of the scale of new physics towards larger scales or even solved the stability problem within a large $\Nf$ approximation for the effective potential. 

Improving the results by taking $1/\Nf$ corrections into account, the space of allowed bare potentials obtained from the mean-field analysis that are compatible with observed IR physics was further constraint. 
At the same time, the extended mean-field analysis offered new mechanisms to shift the scale of new physics towards larger scales.
In particular it turned out that the nonpolynomial structures have to impose strong contributions from the scalar fluctuations. 
This mechanism is remarkable as nonperturbative physics in terms of a strongly coupled Higgs sector is usually associated with the upper Higgs mass bound, here we got a first glance on how these effects might diminish the lower mass bound.

As scalar fluctuations are not considered within the mean-field approximation, a suitable convergence property regarding the $1/\Nf$ expansion cannot be expected. However, we were able to construct one particular family of generalized bare potentials that shows some convergence behavior. For this family an example was given that was able to diminish the lower bound below present consistency bounds obtained from finite-order generalizations of the bare action within the considered toy model \cite{Gies:2013fua,Jakovac:2015kka,Gies:2017zwf}. 
Moreover, we demonstrated how bare potentials can be constructed via reverse engineering such that the effective potential does not suffer from a stability problem and is compatible with observed IR physics.

However, to fully establish these mechanisms a full nonperturbative RG flow is required.
The challenging part of this task is to compute the RG flow with a sufficiently high precision in order to separate the cutoff from the electroweak scale and the scalar potential has to be investigated beyond local approximations to investigate its global properties. Sophisticated solvers based on pseudo-spectral methods have turned out to be useful for this \cite{Borchardt:2015rxa,Borchardt:2016pif,Borchardt:2016kco,Knorr:2016sfs,Knorr:2017yze}. 
We were able to show, that a further diminishing of the lower Higgs mass bound by nonpolynomial bare interactions is possible, if the full flow of the scalar potential is considered for the class of exponential-type bare interactions with an infinite radius of convergence. For this class, the large $\Nf$ expansion captures all relevant effects at least on a qualitative level.

Beyond these technical considerations, this work can be extended in various directions. Even though the Brout-Englert-Higgs effect is much more involved in a theory with local gauge symmetry \cite{tHooft:1979hnm,Osterwalder:1977pc,Banks:1979fi,Frohlich:1980gj,Frohlich:1981yi,Maas:2012tj,Maas:2013aia,Maas:2016ngo,Egger:2017tkd,Maas:2017xzh,Maas:2017pcw}, a generalization of this approximation to the full standard model is, of course, more involved but straightforward. Moreover, we considered only nonpolynomial generalizations of the scalar potential here but also modifications of the kinetic terms might stabilize the effective Higgs potential \cite{Ghoshal:2017egr}. 
Besides solving the stability problem, nonpolynomial structures might also be able to resolve other open problems without introducing new degrees of freedom or symmetries beyond the standard model and offer interesting properties \cite{Matone:2015dya}. 
For instance, the impact of nonpolynomial bare potentials in terms of the building blocks of a resurgent transseries expansion can be investigated to obtain a sufficiently strong first order phase transition in the context of electroweak baryogenesis \cite{Reichert:2017puo}.

In addition, the presented results can be used to constrain the underlying physics of the standard model.
For instance, certain classes of nonpolynomial bare interactions are not compatible with observed IR physics.
In case some theory beyond the standard model generates such a nonpolynomial structure in the bare Higgs potential, it cannot be a viable extension of the standard model.

\section*{Acknowledgements}
It is a pleasure to thank H. Gies and M. Reichert for valuable discussions and in particular H. Gies for comments on the manuscript. 
This work was supported by a postdoc fellowship of the Carl-Zeiss Stiftung.

\bibliography{bibliography}

\end{document}